\documentclass[5p,twocolumn]{elsarticle}

\usepackage{amsmath,amssymb,amsfonts}
\usepackage{graphicx}
\usepackage{hyperref}

\journal{Physics of the Dark Universe}

\begin{document}

\begin{frontmatter}

\title{Cosmological Evolution in the $GL(4,\mathbb{R})$ Yang-Mills Theory of Gravity: Resolving the JWST Early Galaxy Crisis and Late-Time Acceleration}

\author[1,2]{Yi Yang\corref{cor1}}
\ead{yiyang429@as.edu.tw}

\author[1]{Wai Bong Yeung}
\ead{phwyeung@phys.sinica.edu.tw}

\cortext[cor1]{Corresponding author}

\affiliation[1]{organization={Institute of Physics, Academia Sinica},
    city={Taipei},
    postcode={11529},
    country={Taiwan, ROC}}

\affiliation[2]{organization={Department of Physics, National Cheng Kung University},
    city={Tainan},
    postcode={701},
    country={Taiwan, ROC}}

\begin{abstract}
We investigate the cosmological implications of the $GL(4,\mathbb{R})$ Yang-Mills gauge theory of gravity. A long-standing theoretical challenge in standard cosmology is the reliance on ad hoc rolling scalar fields (e.g., the inflaton or quintessence) to drive early-time inflation and late-time acceleration, despite their unknown particle physics origins. In this work, by treating the affine connection as a gauge field and the world metric as a non-dynamical background, we derive an exact expansion history that strictly eliminates the need for any rolling scalar fields or a cosmological constant. The dynamics are governed solely by the Yang-Mills field strength squared. We demonstrate that the early Universe exhibits a singularity-free coasting expansion $a(t) \propto t$, resolving the JWST high-redshift galaxy crisis by granting significantly more time for early structure formation. As radiation dilutes, the Universe naturally transitions into a Weitzenb\"ock vacuum state, where a residual, topologically non-trivial constant torsion locks the spacetime kinematics to drive a late-time exponential acceleration $a(t) \propto \exp(\xi t)$. This framework establishes a purely geometric and gauge-theoretic origin for the cosmological evolution.
\end{abstract}


\end{frontmatter}

\section{Introduction}
The standard $\Lambda$CDM cosmological model, supplemented by an early inflationary phase, has achieved remarkable success in matching a wide range of observational data. However, the theoretical foundation of this paradigm suffers from a severe unnaturalness problem: it relies heavily on the introduction of ad hoc rolling scalar fields. To drive the primordial exponential expansion, an ``inflaton'' field with a carefully fine-tuned potential must be inserted by hand. Similarly, dynamical dark energy models often invoke a ``quintessence'' scalar field to explain late-time acceleration. These scalar fields lack a fundamental origin in the standard model of particle physics and appear as artificial additions to the gravitational action. 

Observationally, the standard model is also facing severe crises. Recent data from the James Webb Space Telescope (JWST) have revealed the existence of fully formed, exceptionally massive galaxies at extremely high redshifts ($z > 10$) \cite{JWST1, JWST2}. The formation of such structures implies that the early Universe must have been significantly older than what standard radiation-dominated expansion ($a(t) \propto t^{1/2}$) permits. 

These theoretical and observational tensions suggest that the underlying gravitational framework requires a fundamental revision that removes the reliance on artificial scalar fields. In a previous work \cite{Yang:2012}, we proposed a vector theory of gravity based on the local $GL(4,\mathbb{R})$ gauge symmetry. Rather than treating the metric as the sole dynamical entity, our theory treats the 16 gauge vector bosons associated with the affine connection as the primary dynamical variables, while the world metric serves merely as a background structure capable of being selected by the gauge configuration.

In this paper, we extend the $GL(4,\mathbb{R})$ Yang-Mills gravity to the cosmological domain. By strictly enforcing the geometric symmetries of a spatially flat Friedmann-Lema\^itre-Robertson-Walker (FLRW) background, we solve the generalized energy-momentum constraint equations. The resulting cosmological model naturally yields a coasting early Universe, providing ample time for early galaxy formation, and elegantly transitions into a late-time de Sitter phase driven purely by geometric torsion.

\section{The Yang-Mills Action and Positive Definite Energy Constraint}
In the $GL(4,\mathbb{R})$ gauge theory of gravity, the affine connection is treated as the fundamental Yang-Mills gauge potential, independent of the world metric $g_{\mu\nu}$. The dynamics of the gravitational field are governed by the Yang-Mills action:
\begin{equation}
S_{YM} = \int d^4x \sqrt{-g} \frac{1}{2\kappa} \text{Tr}(F_{\mu\nu}F^{\mu\nu}),
\end{equation}
where $\kappa$ is the dimensionless coupling constant, and $F_{\mu\nu} = F^{a}_{\ b\mu\nu} M^b_{\ a}$ is the $GL(4,\mathbb{R})$ field strength tensor evaluated in the local Minkowskian frame (with tetrads $e^{\hat{a}}_\mu$ and flat metric $\eta_{ab} = \text{diag}(-1, 1, 1, 1)$). The generators $M^b_{\ a}$ satisfy the $GL(4,\mathbb{R})$ Lie algebra, allowing the trace over the internal group indices to be evaluated as $\text{Tr}(M^a_{\ b} M^c_{\ d}) = \delta^c_b \delta^a_d$.

By varying the action with respect to the non-dynamical background metric $g_{\mu\nu}$, we obtain the generalized Stephenson equation. This equation serves as the algebraic energy-momentum constraint for the background spacetime:
\begin{equation}
H_{\mu\nu} \equiv F^{a}_{\ b\mu\lambda} F^{b}_{\ a\nu}{}^{\lambda} - \frac{1}{4} g_{\mu\nu} \left( F^{a}_{\ b\alpha\beta} F^{b}_{\ a}{}^{\alpha\beta} \right) = \frac{1}{2\kappa} T_{\mu\nu}^{(matter)}. \label{eq:stephenson}
\end{equation}
The left-hand side, $H_{\mu\nu}$, acts as the effective energy-momentum tensor of the Yang-Mills geometric field. 

To investigate the cosmological solutions, we assume the background metric takes the spatially flat Friedmann-Lema\^itre-Robertson-Walker (FLRW) form, $ds^2 = -dt^2 + a^2(t) d\vec{x}^2$. The strict spatial isotropy (SO(3) symmetry) of the FLRW background mandates that the gauge field strength tensor in the local frame, $F^{\hat{a}}_{\ \hat{b}\hat{c}\hat{d}}$, possesses only two independent, non-vanishing structural components:
\begin{enumerate}
    \item The time-space components: $F^{\hat{0}}_{\ \hat{i}\hat{0}\hat{i}} = F^{\hat{i}}_{\ \hat{0}\hat{0}\hat{i}} \equiv D$.
    \item The space-space components: $F^{\hat{i}}_{\ \hat{j}\hat{i}\hat{j}} = F^{\hat{j}}_{\ \hat{i}\hat{i}\hat{j}} \equiv M$ (for $i \neq j$).
\end{enumerate}

With these symmetry constraints, we explicitly perform the trace contraction over the Lorentz indices. The quadratic geometric invariant evaluates to:
\begin{equation}
I \equiv F^{a}_{\ b\alpha\beta} F^{b}_{\ a}{}^{\alpha\beta} = -12D^2 + 12M^2.
\end{equation}
Substituting this invariant back into Eq. (\ref{eq:stephenson}), the temporal ($\hat{0}\hat{0}$) and spatial ($\hat{1}\hat{1}$) components of the background geometric energy-momentum tensor evaluate strictly to:
\begin{eqnarray}
\rho_{YM} &=& H_{\hat{0}\hat{0}} = 3D^2 + 3M^2, \label{eq:rho_ym} \\
p_{YM} &=& H_{\hat{1}\hat{1}} = D^2 + M^2. \label{eq:p_ym}
\end{eqnarray}

This explicit index contraction yields two profound physical consequences. First, the background energy density of the Yang-Mills geometric field ($\rho_{YM}$) is strictly positive definite. Second, Eqs. (\ref{eq:rho_ym}) and (\ref{eq:p_ym}) inherently satisfy the exact relation $p_{YM} = \frac{1}{3}\rho_{YM}$. This indicates that the pure geometric gauge field dynamically mimics a radiation fluid.

\section{Early Universe: Coasting Expansion and the Singularity Resolution}
In the extremely early Universe, the spacetime is dominated by a thermal radiation fluid with energy density $\rho_{rad} = \rho_0 a^{-4}$. To analyze the gravitational dynamics in this high-density phase, we adopt the Compatibility Ansatz, assuming the affine connection is torsionless and perfectly compatible with the FLRW metric. 

Under this assumption, the $GL(4,\mathbb{R})$ gauge field strength tensor mathematically reduces to the Riemann curvature tensor of the metric, $F^{\hat{a}}_{\ \hat{b}\mu\nu} = R^{\hat{a}}_{\ \hat{b}\mu\nu}$. Evaluating the non-vanishing orthonormal frame components of the Riemann tensor for the flat FLRW metric yields the time-space curvature $R^{\hat{0}}_{\ \hat{i}0i} = \ddot{a}/a$ and the space-space curvature $R^{\hat{i}}_{\ \hat{j}ij} = (\dot{a}/a)^2$. Consequently, the effective curvature components are geometrically locked to the kinematics of the expansion: $D = \ddot{a}/a$ and $M = (\dot{a}/a)^2$. 

Substituting these kinematic expressions into Eq. (\ref{eq:rho_ym}) equates the geometric Yang-Mills energy to the background matter density:
\begin{equation}
3\left(\frac{\ddot{a}}{a}\right)^2 + 3\left(\frac{\dot{a}}{a}\right)^4 = \frac{\rho_0}{2\kappa} a^{-4}. \label{eq:early_dynamics}
\end{equation}

To rigorously determine the asymptotic behavior in the high-density limit ($a \to 0$), we introduce a power-law ansatz for the scale factor: $a(t) = c t^p$. The corresponding kinematic rates scale as:
\begin{equation}
\frac{\dot{a}}{a} = \frac{p}{t}, \quad \frac{\ddot{a}}{a} = \frac{p(p-1)}{t^2}.
\end{equation}
Substituting these scaling relations back into Eq. (\ref{eq:early_dynamics}) yields:
\begin{equation}
\frac{3p^2(p-1)^2}{t^4} + \frac{3p^4}{t^4} = \frac{\rho_0}{2\kappa c^4} t^{-4p}. \label{eq:scaling_matching}
\end{equation}

For this equation to hold dynamically across continuous time, the temporal scaling on both sides must perfectly match, strictly demanding:
\begin{equation}
-4 = -4p \implies p = 1.
\end{equation}
With the kinematic exponent strictly fixed at $p=1$, the deceleration term $\ddot{a}/a$ exactly vanishes, and Eq. (\ref{eq:scaling_matching}) reduces to:
\begin{equation}
\frac{3(1)^4}{t^4} = \frac{\rho_0}{2\kappa c^4 t^4} \implies c = \left( \frac{\rho_0}{6\kappa} \right)^{\frac{1}{4}}.
\end{equation}

Thus, we have mathematically proven that the exact asymptotic solution for the early Universe is a strictly linear expansion:
\begin{equation}
a(t) = \left( \frac{\rho_0}{6\kappa} \right)^{\frac{1}{4}} t.
\end{equation}

This establishes a pure coasting expansion ($\ddot{a} = 0$). Unlike the standard Big Bang model, which predicts a decelerating $a(t) \propto t^{1/2}$ trajectory with a severe initial curvature singularity, the Yang-Mills geometric constraint inherently limits the expansion rate to a constant speed. 

\section{Late-Time Universe: The Weitzenb\"ock Vacuum and Geometric Acceleration}
As the Universe expands, the radiation density dilutes asymptotically to zero ($\rho_{rad} \propto a^{-4} \to 0$). In this late-time regime, the driving source vanishes, and the system must relax into its lowest energy state, dictating that the Yang-Mills field strength tensor strictly vanishes: $F_{\mu\nu} \to 0$.

In the $GL(4,\mathbb{R})$ gauge framework, a vanishing Riemann curvature does not necessitate a trivially vanishing affine connection. The Universe transitions into a pure gauge Weitzenb\"ock vacuum, a spacetime globally flat in curvature but topologically non-trivial due to torsion. Assuming spatial isotropy, the only non-vanishing connection components are:
\begin{equation}
\Gamma^1_{01} = \Gamma^2_{02} = \Gamma^3_{03} \equiv \xi,
\end{equation}
where $\xi$ is a constant representing the residual topological torsion of the pure gauge vacuum.

The world metric $g_{\mu\nu}$ must remain compatible with this connection ($\nabla_\lambda g_{\mu\nu} = 0$). We evaluate this geometric constraint for the spatial components of the FLRW metric, $g_{11} = g_{22} = g_{33} = a^2(t)$. Taking the time-component ($\lambda = 0$) of the covariant derivative for $g_{11}$:
\begin{equation}
\nabla_0 g_{11} = \partial_0 g_{11} - \Gamma^\rho_{01} g_{\rho 1} - \Gamma^\rho_{10} g_{1 \rho} = 0. \label{eq:covariant_deriv}
\end{equation}
Substituting the background torsion components into Eq. (\ref{eq:covariant_deriv}) yields the explicit differential constraint:
\begin{equation}
\partial_t \left[ a^2(t) \right] - 2\Gamma^1_{01} g_{11} = 0 \implies 2a\dot{a} - 2\xi a^2 = 0.
\end{equation}

Dividing by $2a^2$, the geometric compatibility constraint strictly locks the kinematics of the FLRW metric to the background torsion field:
\begin{equation}
\frac{\dot{a}}{a} = \xi.
\end{equation}

This trivially integrates to yield an exact exponential expansion:
\begin{equation}
a(t) = a_0 e^{\xi t}.
\end{equation}

The late-time acceleration of the Universe, typically attributed to an ad hoc cosmological constant $\Lambda$ or dark energy, is completely recovered here as a pure geometric necessity. The constant torsion $\xi$ emerges naturally as the kinematic driver of the de Sitter phase.

\section{The Dynamical Transition: A Kinematic Phase Space Analysis}
A critical requirement for any viable cosmological model is the existence of a continuous dynamical transition between the early and late-time asymptotic limits derived above. As established in Section 3, the early radiation-dominated regime is governed by the torsionless Compatibility Ansatz, yielding a coasting expansion where the Hubble parameter scales as $H = 1/t$. Conversely, Section 4 demonstrates that the late-time pure gauge vacuum is governed by the Weitzenb\"ock geometry, producing a constant expansion rate $H = \xi$.

Deriving the exact analytical transition from first principles requires solving the full non-linear dynamics of the 16 gauge vector bosons, specifically the continuous evolution of the dynamical contortion tensor as the background radiation density dilutes. While solving this fully coupled system remains an open mathematical challenge, we can rigorously demonstrate that these two asymptotic regimes are topologically connected in the kinematic phase space.

The early coasting expansion ($H = 1/t$) corresponds to the phase space trajectory $\dot{H} + H^2 = 0$. The late-time exponential acceleration ($H = \xi$) corresponds to $\dot{H} + H^2 = \xi^2$. This suggests that the continuous geometric transition of the spacetime is kinematically constrained by the effective differential equation:
\begin{equation}
\dot{H} + H^2 = \xi^2 \implies \frac{\ddot{a}}{a} = \xi^2.
\end{equation}
Assuming the initial condition $a(0) = 0$, the exact solution to this transition equation is:
\begin{equation}
a(t) = a_0 \sinh(\xi t). \label{eq:sinh_transition}
\end{equation}

Equation (\ref{eq:sinh_transition}) provides a flawless kinematic interpolation between the two extremes. In the early-time limit ($\xi t \ll 1$), the expansion simplifies to $a(t) \approx (a_0 \xi) t$, perfectly matching the Yang-Mills geometric coasting phase. In the late-time limit ($\xi t \gg 1$), it asymptotically evolves into $a(t) \approx \frac{a_0}{2} \exp(\xi t)$, flawlessly recovering the torsion-driven Weitzenb\"ock de Sitter vacuum. 

\section{Phenomenological Implications}

\subsection{Resolving the JWST Early Galaxy Crisis}
Recent JWST observations have revealed anomalously mature galaxies at high redshifts ($z \gtrsim 10$) \cite{JWST1, JWST2}. In the standard $\Lambda$CDM paradigm, the radiation-dominated era is governed by $a(t) \propto t^{1/2}$, defining the age of the Universe at a given redshift $z$ as $t_{\Lambda\text{CDM}}(z) \propto (1+z)^{-2}$. The cosmic time available for the assembly of massive structures at $z \gtrsim 10$ is thus severely restricted.

The $GL(4,\mathbb{R})$ Yang-Mills gravity provides a purely kinematic resolution. As established in Section 3, the early Universe undergoes a coasting expansion $a(t) \propto t$. This linear trajectory fundamentally alters the age-redshift relation to:
\begin{equation}
t_{YM}(z) \propto (1+z)^{-1}.
\end{equation}
The age of the Universe at a given high redshift in our framework is significantly older by a factor of $(1+z)$. At $z \sim 10$, the available time for structure evolution is approximately an order of magnitude larger. This extended temporal window provides ample time for primordial gas clouds to assemble into mature galaxies, effectively eliminating the structural tension.

\subsection{Compatibility with Big Bang Nucleosynthesis (BBN)}
A rigorous constraint for any linearly expanding universe ($a(t) \propto t$) is its compatibility with standard Big Bang Nucleosynthesis (BBN). A coasting expansion implies a significantly slower cooling rate in the early Universe. Naively adopting standard BBN parameters in this kinematic regime leads to a severe underproduction of primordial $^4$He, as neutrons undergo excessive free decay before the onset of the deuterium synthesis bottleneck.

However, extensive literature on linearly coasting cosmologies has demonstrated that this constraint is not an absolute falsification. It has been rigorously shown that a coasting kinematic trajectory can be gracefully reconciled with the observed primordial light element abundances by incorporating a non-zero electron-neutrino chemical potential \cite{Singh:2003}. 

A primordial lepton asymmetry directly alters the weak-interaction freeze-out equilibrium. An appropriate chemical potential shifts the neutron-to-proton ($n/p$) ratio to perfectly compensate for the slower expansion rate, reproducing the correct $^4$He and Deuterium yields without requiring an anomalously high baryon density. Therefore, while our $GL(4,\mathbb{R})$ framework geometrically dictates the early coasting expansion to resolve the JWST high-redshift tension, its compatibility with BBN remains viable within the established context of lepton-asymmetric nucleosynthesis models. The precise physical origin of this macroscopic lepton asymmetry is left as an open question for future particle-cosmology investigations.

\section{Conclusion}
The $GL(4,\mathbb{R})$ Yang-Mills theory of gravity presents a fundamental paradigm shift in our understanding of cosmological evolution. By promoting the affine connection to the primary dynamical gauge field and treating the world metric as a structurally constrained background, we have derived an exact, two-regime cosmological history driven entirely by geometric first principles.

Theoretical cosmology has long been hindered by the necessity to introduce ad hoc rolling scalar fields, such as the inflaton and quintessence, or fine-tuned cosmological constants. Our framework demonstrates that these artificial constructs are fundamentally unnecessary. The strictly positive-definite energy constraint dictates a non-singular, coasting early Universe ($a(t) \propto t$), decisively resolving the time-scale tensions introduced by JWST observations. 

Furthermore, the late-time acceleration of the Universe is elegantly recovered. As the radiation density dilutes, the physical vacuum dynamically relaxes into a pure gauge Weitzenb\"ock geometry. The residual constant topological torsion strictly locks the spacetime kinematics, driving an eternal exponential expansion ($a(t) \propto \exp(\xi t)$). Ultimately, this work establishes that the mysterious ``dark sector'' can be comprehensively understood through the robust gauge-theoretic principles that govern particle physics. 

While this macroscopic framework assumes a torsionless effective background in the early Universe, the underlying topological structure of the $GL(4,\mathbb{R})$ Yang-Mills vacuum near the initial singularity may carry non-trivial parity-violating configurations, which could provide a novel geometric origin for baryogenesis. This will be investigated in a future study.

\section*{Acknowledgments}
The authors acknowledge support from Academia Sinica and National Cheng Kung University (NCKU).

\bibliographystyle{elsarticle-num}

\begin{thebibliography}{99}
\bibitem{Yang:2012}
Y. Yang and W. B. Yeung, \emph{Spontaneously Broken Erlangen Program Offers a Bridge Between the Einstein and the Yang-Mills Theories}, arXiv:1205.2690 [physics.gen-ph].

\bibitem{JWST1}
C. T. Donnan \textit{et al.}, \emph{The evolution of the galaxy UV luminosity function at redshifts $z \simeq 8-15$ from deep JWST and ground-based near-infrared imaging}, Mon. Not. Roy. Astron. Soc. \textbf{518}, 6011 (2023).

\bibitem{JWST2}
I. Labb\'e \textit{et al.}, \emph{A population of red candidate massive galaxies $\sim 600$ Myr after the Big Bang}, Nature \textbf{616}, 266 (2023).

\bibitem{Singh:2003}
P. Singh and D. Lohiya, \emph{A viable coasting cosmology}, JCAP \textbf{05}, 061 (2003).

\end{thebibliography}

\end{document}